\documentclass[conference]{IEEEtran}
\IEEEoverridecommandlockouts
\usepackage{cite}
\usepackage{amsmath,amssymb,amsfonts}
\usepackage{algorithmic}
\usepackage{graphicx}
\usepackage{textcomp}
\usepackage{xcolor}
\usepackage{multirow}
\usepackage{booktabs}
\usepackage{hyperref}
\usepackage{authblk}
\usepackage{subcaption}
\hypersetup{
    colorlinks=true,
    linkcolor=blue,
    filecolor=magenta,      
    urlcolor=cyan,
    pdfpagemode=FullScreen,
}

\def\BibTeX{{\rm B\kern-.05em{\sc i\kern-.025em b}\kern-.08em
    T\kern-.1667em\lower.7ex\hbox{E}\kern-.125emX}}
\begin{document}

\title{\huge{
iELAS: An ELAS-Based Energy-Efficient Accelerator for Real-Time Stereo Matching on FPGA Platform} 
}

\author[$1$]{Tian~Gao\textsuperscript{*}}
\author[$2$]{Zishen~Wan\textsuperscript{*}\thanks{\textsuperscript{*}These authors contributed equally to this work.}}
\author[$1$]{Yuyang~Zhang}
\author[$3$]{Bo~Yu}
\author[$1$]{Yanjun~Zhang}
\author[$3$]{Shaoshan Liu} 
\author[$2$]{Arijit~Raychowdhury\thanks{\textit{IEEE International Conference on Artificial Intelligence Circuits and Systems (AICAS)}, June 6-9, 2021, Virtual.}}

\affil[$1$]{\small School of Information and Electronics, Beijing Institute of Technology, Beijing, 100081, China}
\affil[$2$]{\small School of Electrical and Computer Engineering, Georgia Institue of Technology, Atlanta, GA 30332, USA}
\affil[$3$]{\small PerceptIn Inc, Fremont, CA 94539, USA}

\maketitle

\begin{abstract}
Stereo matching is a critical task for robot navigation and autonomous vehicles, providing the depth estimation of surroundings. Among all stereo matching algorithms, Efficient Large-scale Stereo (ELAS) offers one of the best tradeoffs between efficiency and accuracy. However, due to the inherent iterative process and unpredictable memory access pattern, ELAS can only run at 1.5-3~\textit{fps} on high-end CPUs and difficult to achieve real-time performance on low-power platforms. In this paper, we propose an energy-efficient architecture for real-time ELAS-based stereo matching on FPGA platform. Moreover, the original computational-intensive and irregular triangulation module is reformed in a regular manner with points interpolation, which is much more hardware-friendly. Optimizations, including memory management, parallelism, and pipelining, are further utilized to reduce memory footprint and improve throughput. Compared with Intel i7 CPU and the state-of-the-art CPU+FPGA implementation, our FPGA realization achieves up to 38.4$\times$ and 3.32$\times$ frame rate improvement, and up to 27.1$\times$ and 1.13$\times$ energy efficiency improvement, respectively.


\end{abstract}

\section{Introduction}
\label{sec:intro}
Stereo vision is a critical component in many computer vision applications, such as robot navigation, autonomous vehicles, augmented reality, and gesture recognition~\cite{wan2020survey}. Especially, stereo vision provides the availability of depth estimation to abstract 3D scene structure from 2D images captured by two cameras. Accurate depth information is crucial for the safety of autonomous systems and serves as the prerequisite for following localization and motion planning~\cite{wan2021energy}.

Efficient Large-Scale Stereo (ELAS)~\cite{geiger2010efficient} has recently attracted increasing attention and becomes one of the most accurate stereo matching algorithms. It overcomes the downsides of local matching methods that perform poorly on ambiguous surfaces and the downsides of global matching methods that require extensive computational efforts and high memory capacities. ELAS enables dense matching with small aggregation windows by reducing ambiguities on correspondences.

However, it has been shown that ELAS can only reach 1.5-3~\textit{fps} speed on a high-end CPU~\cite{geiger2010efficient}. Some intermediate steps in ELAS algorithm, such as the support points triangulation, involve a significant number of sequential, iterative, and condition processes with unpredictable memory accesses, posing great challenges for implementing ELAS on resource-constrained edge devices and limiting its application in latency and power-sensitive applications and autonomous cyber-physical systems~\cite{krishnan2020sky,krishnan2021machine}.

Some prior efforts have been made to accelerate ELAS on low-power platforms. \cite{rahnama2018real} implements ELAS on an embedded FPGA-CPU SoC and achieves 17.3~\textit{fps} on KITTI dataset~\cite{geiger2013vision} under 3~\textit{W} power. \cite{rahnama2019real} further introduces an ELAS-SGM (Semi Global Matching) combination approach and evaluates it on an FPGA-CPU platform, achieving 50~\textit{fps} and 4.5~\textit{W} power consumption on the KITTI dataset. However, both \cite{rahnama2018real} and \cite{rahnama2019real} accelerate only part of ELAS modules on FPGA and offload the triangulation and grid vector extraction computation on CPU during processing, which brings in higher power consumption and performance degradation.

In this paper, \textit{iELAS} is proposed as a fully FPGA-based architecture of ELAS stereo vision system. The most time-consuming and iterative triangulation procedures are accelerated on FPGA by introducing an extra support points interpolation module, making it much more hardware-friendly (Sec.~\ref{subsec:optimized_ELAS}). By utilizing parallelisms, pipelining, ping-pong storage mechanism, and several memory management techniques, all modules of \textit{iELAS} are fully accelerated on an FPGA platform (Sec.~\ref{sec:arch}). Our design achieves up to 38.4$\times$ and 3.32$\times$ speedup over Intel i7 CPU and the state-of-the-art FPGA+Arm ELAS implementation~\cite{rahnama2018real}, with 27.1$\times$ and 1.13$\times$ lower power consumption, respectively (Sec.~\ref{sec:eval}).

The main contributions of this paper are as follows:
\begin{itemize}
    \item A novel ELAS-based stereo vision hardware accelerator is proposed for real-time applications on energy-efficient FPGA platforms.
    
    \item An interpolated optimized support point extraction pattern is utilized to make ELAS algorithm much more hardware-friendly.
    
    \item Optimizations including memory management, parallelism, and pipelining are further exploited to save hardware resource consumption and improve throughput.
\end{itemize}

\section{Algorithm Framework}
\label{sec:algo}
In this section, we present the original ELAS algorithm flow (Sec.~\ref{subsec:original_ELAS}), and then propose an interpolated optimized ELAS algorithm to make it much more hardware-friendly (Sec.~\ref{subsec:optimized_ELAS}).
\subsection{Original ELAS Algorithm}
\label{subsec:original_ELAS}

ELAS algorithm is inspired by the observation that despite many stereo correspondences being highly ambiguous, some can be robustly matched. It first establishes a prior over the disparity space by forming a triangulation on a set of correspondences whose estimation is simpler and comes with a higher degree of confidence. These correspondences provide a rough approximation of the scene geometry and guide the dense matching stage. ELAS is attractive since the slanted plane prior can be very efficiently implemented, and the dense depth estimation is fully decomposable over all pixels.

\begin{figure}[t!]
    \begin{subfigure}{.98\linewidth}
        \includegraphics[width=\textwidth]{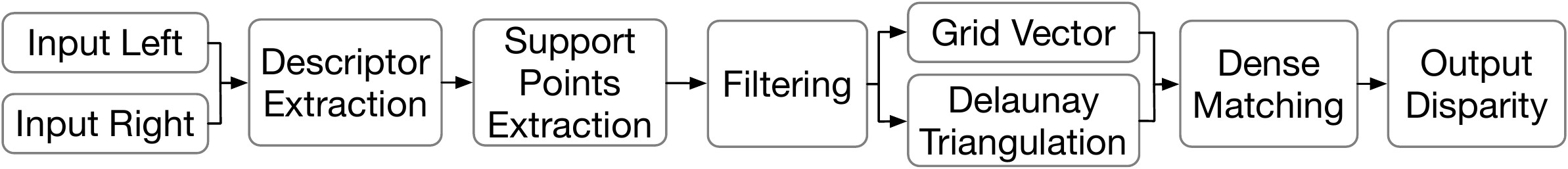}
        \caption{The original ELAS algorithm flow.}
        \label{fig:orginal_ELAS}
    \end{subfigure}
    \vspace{5pt}
    \begin{subfigure}{.98\linewidth}
        \includegraphics[width=\textwidth]{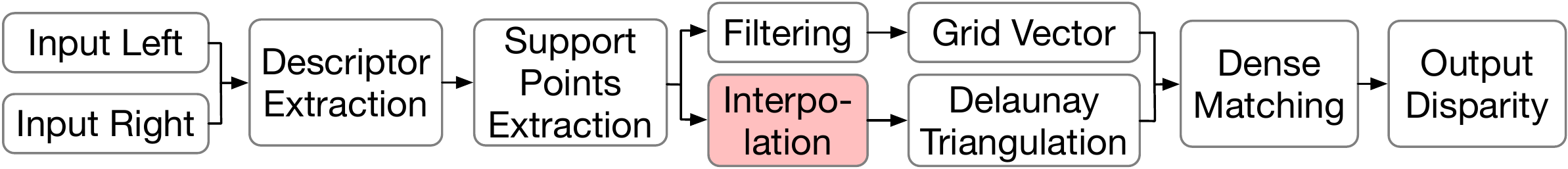}
        \caption{The interpolated optimized ELAS algorithm flow.}
        \label{fig:optimized_ELAS}
    \end{subfigure}
    \caption{ELAS algorithm flow.}
    \label{fig:ELAS}
    \vspace{-10pt}
\end{figure}

Fig.~\ref{fig:orginal_ELAS} demonstrates the overview of the ELAS algorithm. The detailed steps are described as follows:

\textbf{Descriptor Extraction.} 
The input stereo image pairs (left and right images) first pass through Sobel filters over the horizontal and vertical gradients to extract the descriptor information of each pixel. 

\textbf{Support Point Extraction.} 
A set of sparse but confident correspondences (support points) is calculated using descriptor information over the full disparity range.

\textbf{Filtering.} 
By comparing support point values to neighbors within a window region, two types of obtained support points are removed.
The implausible values that are inconsistent would respectively corrupt the representation are filtered out.
The redundant values that are identical to neighbors in the same row or column and unnecessarily complicate the coarse representation are also removed.

\textbf{Disparity Computation.}
The filtered support points are then used to guide the dense stereo matching in two separate ways. First, a slanted plane prior is constructed to approximate coarse scene geometry by using Delaunay triangulation. Second, a grid vector is created by pooling support points within a sub-region to make stereo matching more reliable.

\textbf{Post-processing.} 
ELAS uses post-processing techniques, such as gap interpolation, median filtering, to invalidate occluded pixels and further smooth the images.

\subsection{Interpolated Optimized ELAS Algorithm}
\label{subsec:optimized_ELAS}
In the original ELAS algorithm, the Delaunay triangulation module involves a significant number of speculations and random memory accesses, posing great challenges in FPGA acceleration. \cite{rahnama2018real} chooses to offload triangulation computation to the CPU core, bringing higher power consumption and latency. To make Delaunay triangulation computing more hardware-friendly, we put forward a special way to interpolate support points and enable regular Delaunay triangulation procedures, which can be fully accelerated in the FPGA platform.

The interpolated ELAS algorithm flow is shown in Fig.~\ref{fig:optimized_ELAS}. After extracting support points, we add an interpolation module to derive a set of newly support points with fixed numbers and coordinations, which significantly facilitates the construction of slanted planes. The interpolation is performed using the disparity value of the support points within the neighborhood to fill the vacant positions. The detailed steps are as follows:

\textbf{Horizontal Interpolation.} For the position $s$ to be interpolated, first search the support points in the horizontal direction within $(s-s_\delta,s+s_\delta)$ window. If there are support points $(P_L, P_R)$ lying on both sides and their disparity values $|D_{P_L}-D_{P_R}|\leq\epsilon$, then we use the mean of $(D_{P_L},D_{P_R})$ to interpolate. If $|D_{P_L}-D_{P_R}|>\epsilon$, then $min(D_{P_L},D_{P_R})$ will be chosen for interpolation.

\textbf{Vertical Interpolation.} If no support point pair $(P_L, P_R)$ is found in the horizontal direction, then search in the vertical direction to find $(P_T, P_B)$ and perform interpolation using the same method as step 1.

\textbf{Constant Interpolation.} If no support point pairs are found in both horizontal and vertical directions, then fill a constant disparity value $C$ in the position $s$.

\begin{figure}[t!]
        \centering\includegraphics[width=.75\columnwidth]{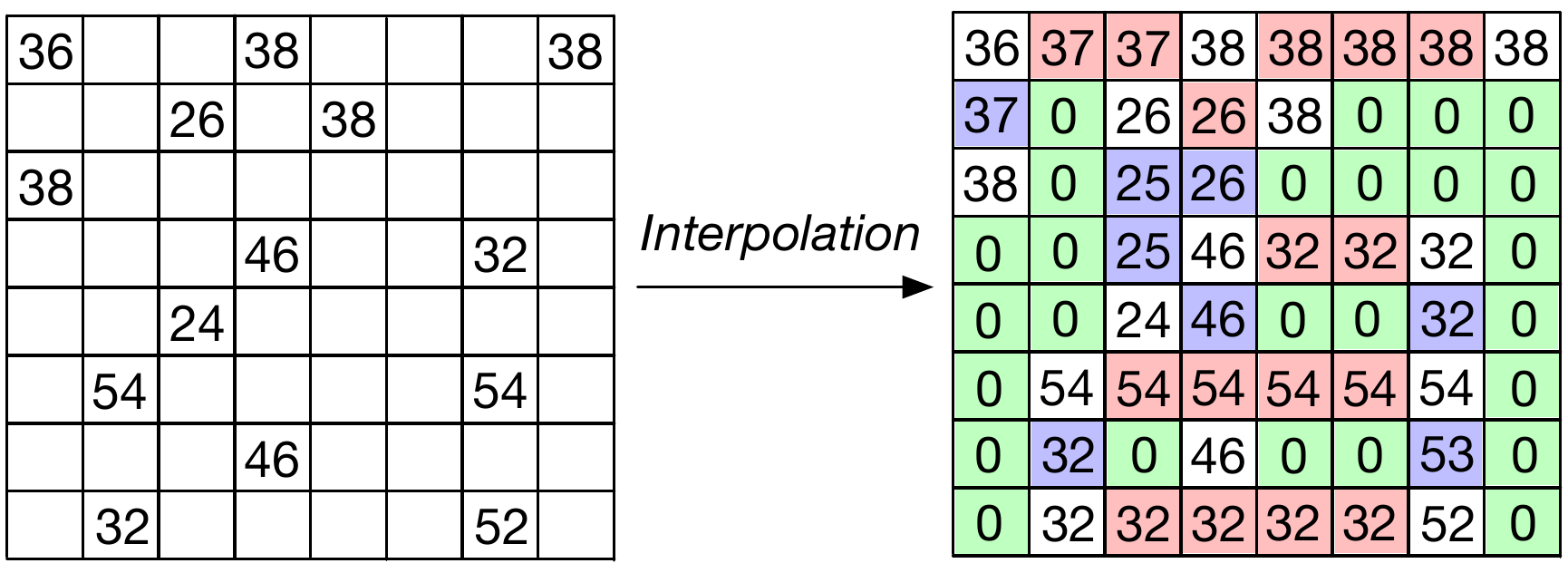}
        \caption{Exampled support points interpolation ($s_\delta=5$, $\epsilon=3$, $C=0$). Red represents horizontal interpolation, blue represents vertical interpolation, green represents constant interpolation.}
        \label{fig:sp}
\end{figure}

Following this procedure, we present an example of support points interpolation in fig.~\ref{fig:sp}, where $s_\delta=5$, $\epsilon=3$, $C=0$.
\begin{figure}[t!]
\centering
    \begin{subfigure}{.4\linewidth}
        \includegraphics[width=\textwidth]{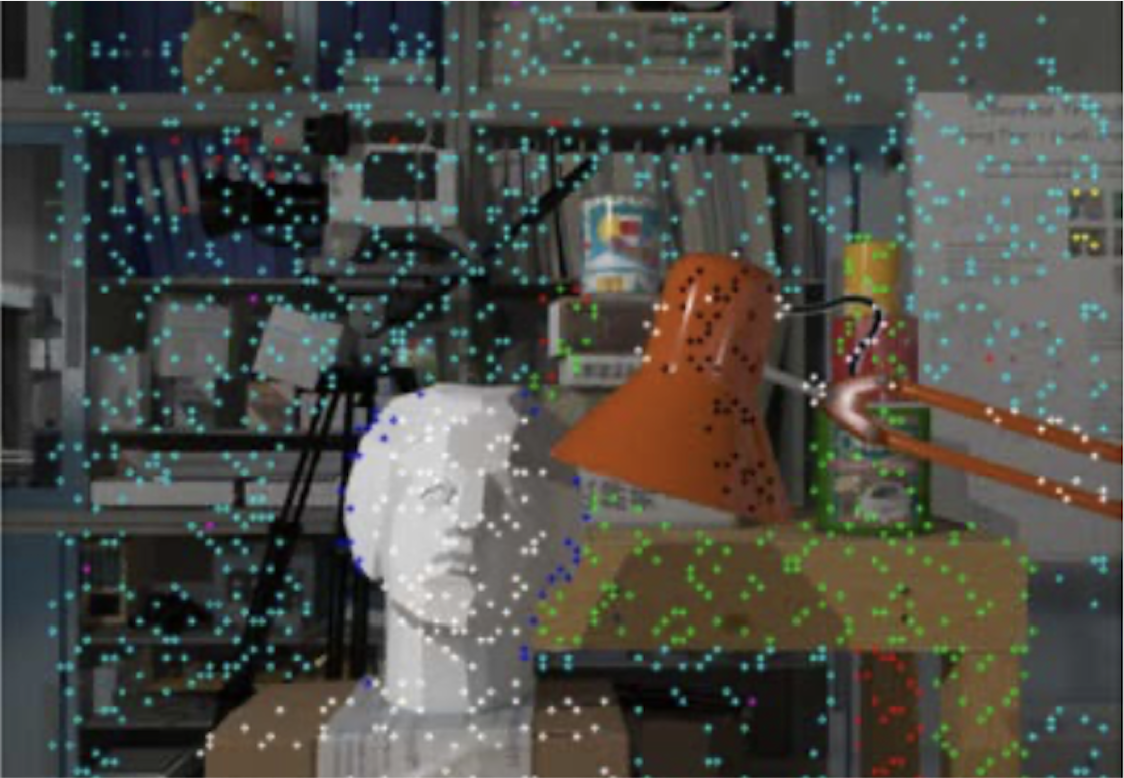}
        \caption{Original ELAS.}
        \label{fig:orginal_sp}
    \end{subfigure}
    \hspace{5pt}
    \begin{subfigure}{.4\linewidth}
        \includegraphics[width=\textwidth]{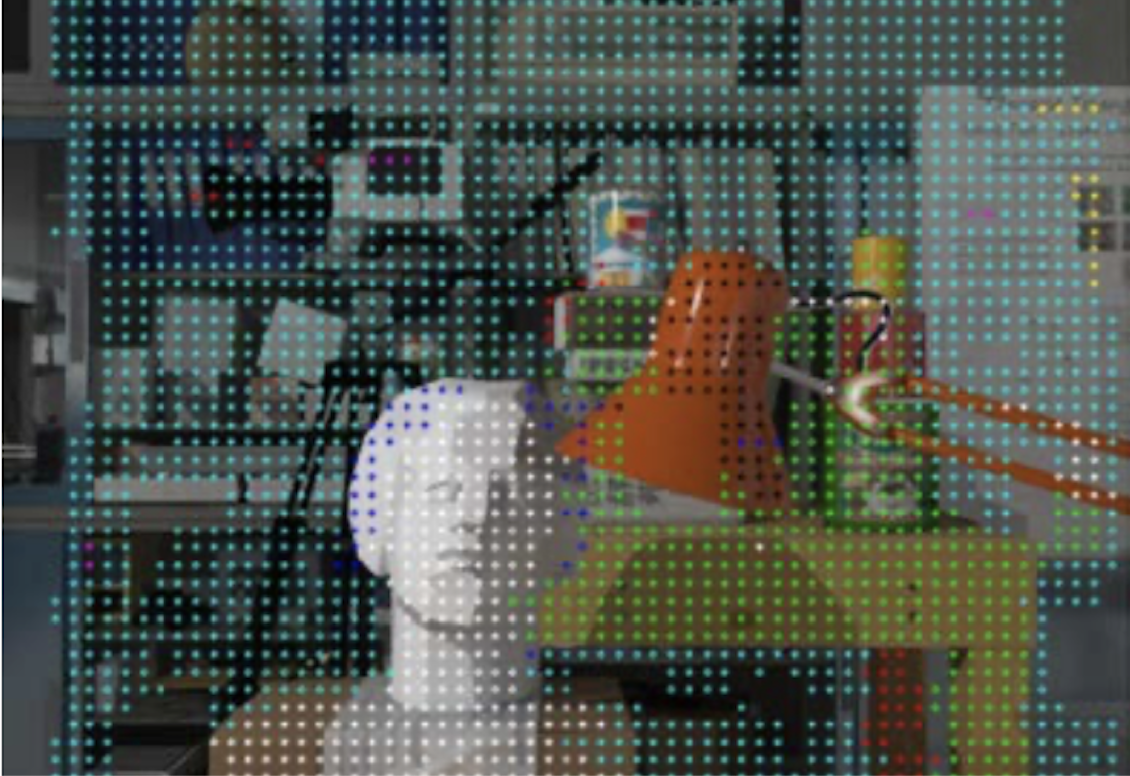}
        \caption{Interpolated ELAS.}
        \label{fig:optimized_sp}
    \end{subfigure}
    \caption{Support point visualization.}
    \label{fig:interpolation}
\end{figure}

\begin{table}[t!]
\caption{Error Evaluation of Interpolated ELAS Algorithm.}
\resizebox{\columnwidth}{!}{%
\begin{tabular}{ccccc}
\toprule[0.6pt]
\textbf{Dataset}                                                                         & \textbf{Lightning}   & \textbf{Error (orig.)}  & \textbf{Error (inter.)} & \textbf{Improvement} \\
\hline
\multirow{4}{*}{\begin{tabular}[c]{@{}c@{}}New\\ Tsukuba\\ Stereo\end{tabular}} & Daylight    & 1.84\%         & 1.28\% & 0.56\%  \\
                                                                                & Flashlight  & 1.03\%         & 0.34\%    & 0.69\%      \\
                                                                                & Fluorescent & 3.15\%         & 1.83\%     & 1.32\%     \\
                                                                                & Lamps       & 5.13\%         & 2.24\% & 2.89\%        
\\              \bottomrule[0.6pt]      
\end{tabular}
\label{tab:accuracy}
}
\vspace{-5pt}
\end{table}

The accuracy of the proposed interpolated ELAS is evaluated on the New Tsukuba Stereo Dataset~\cite{martull2012realistic}. Fig.~\ref{fig:orginal_sp} visualize the support points obtained from the original ELAS algorithm, where different colors represent the ranges of different disparity values.
Fig.~\ref{fig:optimized_sp} demonstrates the results after interpolation. It is well observed that the disparity ranges of interpolated support points align with the original set, but the coordinates of the new set of support points are much more regular. This regular pattern will significantly facilitate the following Delaunay triangulation procedure. 

We further evaluate the disparity error across the whole image as:
\begin{equation}
    Error = \frac{1}{N}\sum\frac{|D_{interpolated}-D_{real}|}{D_{real}}
\end{equation}
where $D_{interpolated}$ is the calculated disparity value and $D_{real}$ is the groundtruth disparity value.

Tab.~\ref{tab:accuracy} demonstrates the matching error rate under different lightning conditions.We can observe that the accuracy of our proposed interpolated ELAS algorithm surpasses the traditional ELAS algorithm in all scenarios.

\section{Hardware Architecture}
\label{sec:arch}
In this section, we provide an overview of the proposed \textsl{iELAS} accelerator (Sec.~\ref{subsec:arch_overview}) as well as the detailed implementations of each module (Sec.~\ref{subsec:arch_module}). Several leveraged optimization techniques are presented in Sec.~\ref{subsec:memory}.
\subsection{Hardware Architecture Overview}
\label{subsec:arch_overview}
\begin{figure}[t!]
        \centering\includegraphics[width=\columnwidth]{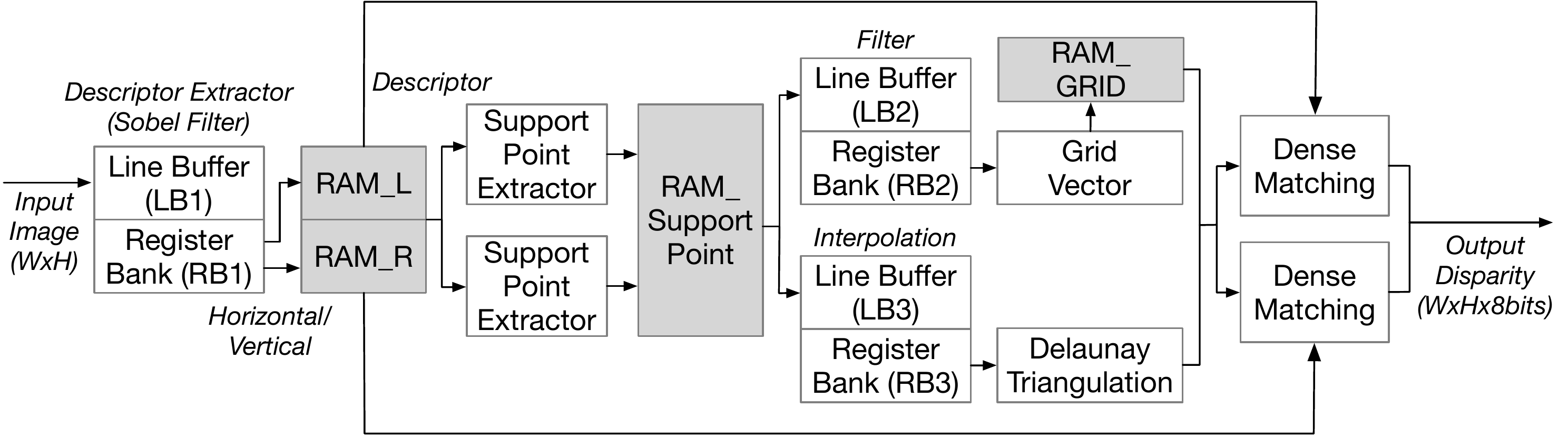}
        \caption{Overview of proposed ELAS accelerator.}
        \label{fig:arch_overview}
        \vspace{-5pt}
\end{figure}
The overall architecture of the proposed ELAS-based accelerator, \textit{iELAS}, is demonstrated in Fig~\ref{fig:arch_overview}. It is fully accelerated on programmable logic of FPGA. The descriptor extractor is implemented to accelerate feature information extraction from input stereo image pairs. Support point extractor and dense matching blocks are responsible for disparity value calculation by leveraging interpolated support points. Several design traits, such as memory management, parallelism, and pipelining are proposed to improve throughput and save hardware resources.

\subsection{Hardware Architecture of Each Compute Module}
\label{subsec:arch_module}
The detailed hardware architecture implementation of each compute module is described as follows.
\begin{figure}[t!]
        \centering\includegraphics[width=.55\columnwidth]{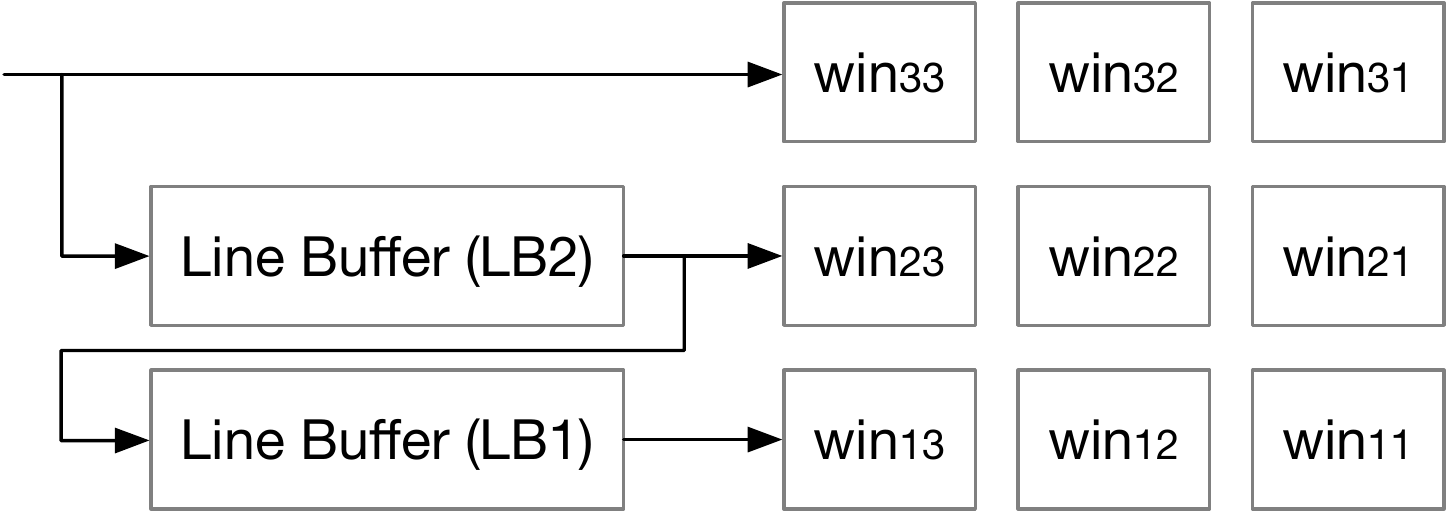}
        \caption{Sobel filter architecture.}
        \label{fig:descriptor}
        \vspace{-10pt}
\end{figure}

\textbf{Descriptor Extractor.}
This module is responsible for obtaining image pair's descriptor information for further support points extraction and dense matching use.
The input image pair is processed by 3$\times$3 Sobel filter in both horizontal and vertical direction, and the pre-processed results are stored in Block RAM (BRAM) for further support points extraction and dense matching use. The Sobel filter is operated as:
\begin{equation}
    F_{3\times 3} = 
\begin{bmatrix}
1 & 0 & -1 \\ 
2 & 0 & -2\\ 
1 & 0 & -1
\end{bmatrix}\begin{bmatrix}
win_{11} & win_{12} & win_{13}\\ 
win_{21} & win_{22} & win_{23}\\ 
win_{31} & win_{32} & win_{33}
\end{bmatrix}
\end{equation}
where $win_{ij}$ is the pixel value of input images.

The descriptor extractor module is mainly composed of line buffers and register banks, as shown in Fig.~\ref{fig:descriptor}. Since the image is fed row-wisely, the length of line buffers is the same as the image width. When the first pixel reaches $win_{22}$ position, the extraction process will be triggered and the descriptor of current patch will be calculated. By performing Sober filtering on $w\times h$ image in this way, all descriptors will be extracted in $(w+1)$ clock cycles, enabling real-time processing.


\begin{figure}[t!]
        \centering\includegraphics[width=.65\columnwidth]{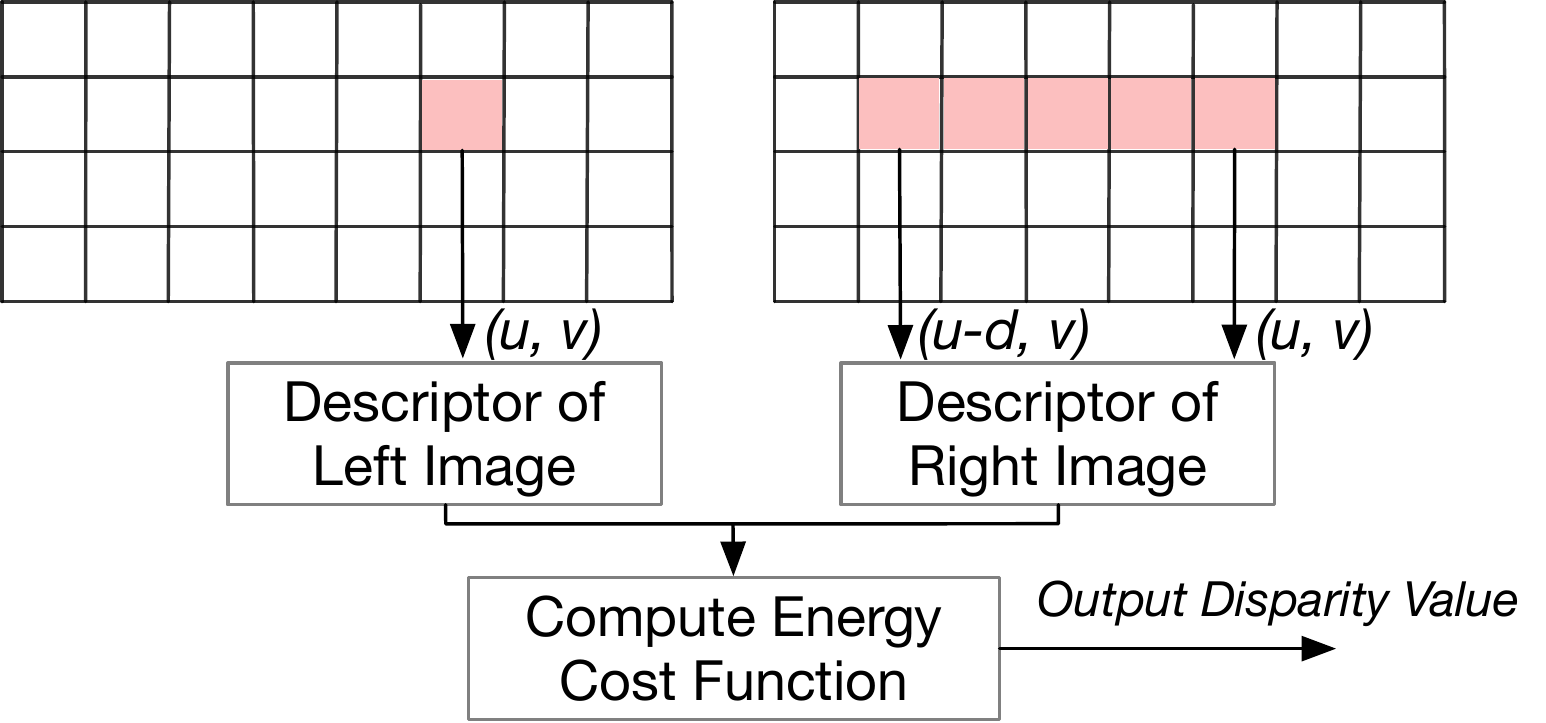}
        \caption{Support point extractor.}
        \label{fig:arch_matching}
        \vspace{-10pt}
\end{figure}


\textbf{Support Point Extractor.} This module extracts a sparse set of support points based on obtained descriptors in any 5$\times$5 window and generates disparity values with their coordinates. As shown in Fig.~\ref{fig:arch_matching}, a specific pixel descriptor $(u,v)$ is selected in left image, and the energy cost function is calculated between $(u,v)$ and each neighbor descriptors in right image. Next, a point in right image is selected and the same process is repeated. The pair with the minimum energy cost value will be considered as the matching pair, and the disparity value is computed accordingly. The results are stored in BRAM.

\textbf{Filtering.} This module removes both implausible and redundant values from disparity values. It consists of line buffers and register banks. The filtered results are sent to the grid vector module. 

\textbf{Grid Vector.} The grid vector module takes the filtered disparity results as input. It aggregates them to limit the disparity values evaluated in the dense matching stage, which can improve the robustness of the system. The grid vector results are stored in BRAM for dense matching use.

\textbf{Interpolator.} This module performs disparity value interpolation horizontally and vertically within the neighbor of extracted support point, significantly facilitating the Delaunay triangulation implementation.

\textbf{Delaunay Triangulator.} It takes the interpolated support points as input and constructs a mesh that can approximate coarse scene geometry and guide the dense matching stage.

\textbf{Dense matching.} The dense matching module takes descriptors, grid vector, and constructed mesh as inputs, and calculates the disparity values of each pixel in input image pairs. The results are represented in 8-bit.

\subsection{Key Accelerator Design Traits}
\label{subsec:memory}

\begin{figure}[t!]
        \centering\includegraphics[width=\columnwidth]{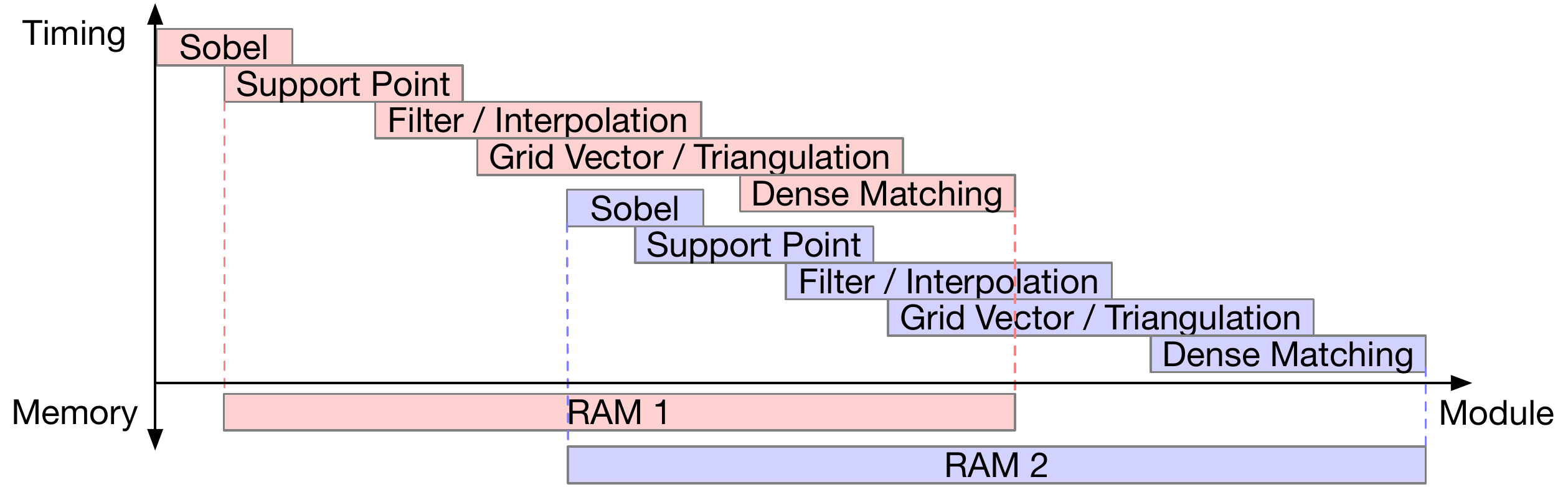}
        \caption{Ping-pong memory management scheme.}
        \label{fig:arch_pingpong}
        \vspace{-10pt}
\end{figure}

To reduce resources consumption and improve throughput, we use several optimization techniques in the \textsl{iELAS} design.


\textbf{BRAM Saving.}
Original ELAS design requires high memory capacities. The descriptor of each pixel is concatenated to 128-bit, and 2400 BRAMs are needed for an image pair. To reduce memory consumption, we propose to directly store 8-bit intermediate results after Sobel filter, making descriptor concatenation task complete during support points extraction, which can achieve around 8$\times$ memory consumption reduction.


\textbf{Parallelism.} Considering both support point and descriptor extractions use 5$\times$5 window, we choose to store every five rows in one BRAM, significantly facilitating the parallelism of support points extraction and dense matching. In this way, both modules can be finished within 17~\textit{ms}, whereas they need 271.6~\textit{ms} and 374.4~\textit{ms} in the original design, respectively.

\textbf{Ping-pong Storage Mechanism.} To improve the throughput of \textsl{iELAS}, we adopt a ping-pong storage mechanism (Fig.~\ref{fig:arch_pingpong}). During processing, frame $i+1$ will arrive before the process of frame $i$ is finished in BRAM 1. To avoid potential data loss of frame $i$, we store the information of frame $i+1$ in BRAM 2 
during the procedure. When frame $i+2$ arrives, BRAM 1 has been released and ready for repeated use. This ping-pong mechanism can improve system's throughput by almost 2$\times$.

\textbf{Grid Vector Optimization.} Original ELAS design stores all 256 disparity values for grid vector computation. However, we notice that most of the support points are removed in the filtering module and cannot be fully utilized by the grid vector. By analyzing the accuracy and resource consumption under the different number of stored support points, we choose to store 20 disparity values of each pixel for grid vector, which can greatly save memory capacity without accuracy degradation.

\section{Evaluation Results}
\label{sec:eval}
This section evaluates the hardware resource utilization (Sec.~\ref{subsec:setup}) and stereo matching accuracy (Sec.~\ref{subsec:result_acc}) of proposed \textsl{iELAS} accelerator, and demonstrates its advantages in performance and energy efficiency (Sec.~\ref{subsec:performance}). 

\subsection{Experiment setup}
\label{subsec:setup}

\textbf{Hardware Platform and Dataset.}
The proposed \textit{iELAS} accelerator is fully implemented on Virtex-7 VC707 FPGA, and evaluated on the New Tsukuba Stereo dataset and KITTI dataset with 640$\times$480 and 1242$\times$375 resolution, respectively.

\textbf{Resource Utilization.}
Tab.~\ref{tab:resource} demonstrates the resource utilization of the proposed \textsl{iELAS} under two datasets.
The FPGA device has 304K LUTs, 607K Flip-Flops, and 1030 BRAMs in total. Overall, the hardware architecture consumes 37.62\% LUT, 9.62\% Flip-Flop, and 48.70\% BRAM on the New Tsukuba Stereo dataset and 47.71\% LUT, 12.99\% Flip-Flop, and 45.56\% BRAM on KITTI dataset.

\begin{table}[t!]
\caption{The FPGA resources utilization of \textit{iELAS}}
\resizebox{\columnwidth}{!}{%
\begin{tabular}{cc|cc|cc}
\toprule[0.5pt]
              & \multirow{2}{*}{\textbf{\begin{tabular}[c]{@{}c@{}}Total\\ Resouces\end{tabular}}} &  \multicolumn{2}{c|}{\textbf{640$\times$480}} & \multicolumn{2}{c}{\textbf{1242$\times$375}} \\ \cline{3-6}
              &                                                                                    & \textbf{Consumption}  & \textbf{Percentage} & \textbf{Consumption}  & \textbf{Percentage}  \\ \hline
\textbf{LUT}  & 303600                                                                             & 114214                & 37.62\%             & 144848                & 47.71\%              \\
\textbf{FF}   & 607200                                                                             & 58412                 & 9.62\%              & 74625                 & 12.99\%              \\
\textbf{BRAM} & 1030                                                                               & 501.5                 & 48.70\%             & 469.5                 & 45.56\%      \\          
\bottomrule[0.5pt]
\end{tabular}
\label{tab:resource}
}
\vspace{-5pt}
\end{table}

\subsection{Accuracy Analysis}
\label{subsec:result_acc}
\begin{table}[]
\centering
\caption{Matching error evaluation.}
\resizebox{\columnwidth}{!}{%
\begin{tabular}{cccc}
\toprule[0.6pt]
\textbf{Dataset (Resolution)}     & \textbf{i7 CPU} & \textbf{FPGA+ARM}~\cite{rahnama2018real} & \textbf{FPGA (Ours)} \\ \hline
\textbf{New Tsukuba (640$\times$480)} & 6.4\%        & 6.8\%             & 7.7\%                   \\
\textbf{KITTI (1242$\times$375)}       & 17.9\%       & 18.4\%            & 19.8\%                  
\\              \bottomrule[0.6pt]      
\end{tabular}
\label{tab:result_accuracy}
}
\vspace{-5pt}
\end{table}

The accuracy of \textsl{iELAS} is measured by matching error which means the number of estimated disparities differing from ground truth (same method as~\cite{rahnama2018real}). As shown in Tab.~\ref{tab:result_accuracy}, comparing with software implementation (i7 CPU) and the state-of-the-art SoC implementation (FPGA+ARM)~\cite{rahnama2018real} on two datasets, \textsl{iELAS} can maintain similar matching accuracy after support points interpolation (under $s_\delta=50$, $\epsilon=15$, $C=60$).



\subsection{Performance Evaluation}
\label{subsec:performance}

\begin{table}[t!]
\caption{Performance and Power Evaluation.}
\resizebox{\columnwidth}{!}{%
\begin{tabular}{ccccc}
\toprule[0.6pt]
                       & \textbf{\begin{tabular}[c]{@{}c@{}}Perform. (\textit{fps})\end{tabular}} & \textbf{\begin{tabular}[c]{@{}c@{}}Perform. (\%)\end{tabular}} & \textbf{\begin{tabular}[c]{@{}c@{}}Power (\textit{W})\end{tabular}} & \textbf{\begin{tabular}[c]{@{}c@{}}Power (\%)\end{tabular}} \\ \hline
\multicolumn{5}{c}{New Tsukuba Stereo Dataset (640$\times$480)}                                                                                                                                                                                                                                                      \\ \hline
\textbf{FPGA (Ours)}           & 57.6                                                                  & -                                                                   & 2.40                                                         & -                                                             \\
\textbf{ARM+FPGA~\cite{rahnama2018real}}       & 17.6                                                                   & 30.6\%                                                             &2.70                                                         & 1.13$\times$                                                         \\
\textbf{Intel i7 Core}     &                1.5-3                                                     & 2.6\%-5.2\%                                                              & 65                                                        &  27.1$\times$                                                       \\ \hline
\multicolumn{5}{c}{KITTI Dataset (1242$\times$375)}                                                                                                                                                                                                                                                     \\ \hline
\textbf{FPGA (Ours)}           & 57.5                                                                & -                                                                   & 2.45                                                         & -                                                             \\
\textbf{ARM+FPGA~\cite{rahnama2018real}}    & 17.3                                                                    & 30.1\%                                                             & 2.61                                                             & 1.07$\times$                                                         \\
\textbf{Intel i7 Core} & 1.5-3                                                                   & 2.6\%-5.2\%                                                             & 65                                                            & 26.5$\times$    \\                                                  
\bottomrule[0.6pt]
\end{tabular}
\label{tab:performance}
}
\vspace{-5pt}
\end{table}
\textbf{CPU Comparison.} 
Tab.~\ref{tab:performance} compares the frame rate and power of \textsl{iELAS} and Intel i7 CPU. Compared with CPU, \textsl{iELAS} achieves 38.4$\times$ and 38.3$\times$ speedup with 27.1$\times$ and 26.5$\times$ energy efficiency improvement on New Tsukuba Stereo dataset and KITTI dataset, respectively.

\textbf{Existing Accelerator Comparison.} Tab.~\ref{tab:performance} also compares \textsl{iELAS} with FPGA+Arm implementation~\cite{rahnama2018real}. Compared with~\cite{rahnama2018real}, \textsl{iELAS} raises the performance by 3.27$\times$ and 3.32$\times$, and improves energy efficiency by 1.13$\times$ and 1.07$\times$ for two datasets, respectively. This benefits from the fact that we achieve FPGA acceleration for all compute modules by leveraging support point interpolation, while \cite{rahnama2018real} offloads Grid Vector and Delaunay Triangulation modules computation on Arm CPU and other modules on FPGA.



\section{Conclusion}
\label{sec:conclusion}
In this paper, an ELAS-based stereo vision system, \textsl{iELAS}, is proposed for real-time and energy-efficient applications and fully implemented on FPGA. The ELAS algorithm is reformulated as a regular pattern with support points interpolation for hardware-friendly implementation. All modules in ELAS are accelerated on FPGA to reduce the latency significantly. The \textsl{iELAS} is also designed in a ping-pong storage pattern with other memory management techniques to further reduce memory footprint and improve the throughput. The evaluation results on the KITTI dataset have shown \textsl{iELAS} could achieve up to 3.32$\times$ and 38.4$\times$ speedup in frame rate, and 1.13$\times$ and 27.1$\times$ improvement in energy efficiency when compared to the state-of-the-art Arm+FPGA and Intel i7 CPU, respectively.
\vspace{-2pt}
\section*{Acknowledgements}
This work was supported in part by C-BRIC, one of six centers in JUMP, a Semiconductor Research Corporation (SRC) program sponsored by DARPA.

\bibliographystyle{ieeetr}
\bibliography{refs}

\end{document}